\documentclass[9pt,a4paper,reqno,twoside,psamsfonts]{amsart}
\usepackage[latin1]{inputenc} \usepackage[T1]{fontenc}
\usepackage{textcomp} \usepackage[british]{babel}
\usepackage{a4wide}
\usepackage{subfigure}
\usepackage{newcent} 
\newif\ifpdf \pdffalse 
  \usepackage[dvips]{graphicx} 
  \usepackage{AUS-math}
\newcommand{\denota}{A}
\newcommand{\denotb}{B}
\newcommand{\denotc}{\bullet}
\newcommand{\renda}[1]{\emox{#1^{\denota}}}
\newcommand{\rendb}[1]{\emox{#1^{\denotb}}}
\newcommand{\rendc}[1]{\emox{#1^{\denotc}}}
\newcommand{\sat}{\emox{s}}
\newcommand{\saa}{\renda{\sat}}
\newcommand{\sbb}{\rendb{\sat}}
\newcommand{\scc}{\rendc{\sat}}
\newcommand{\util}[1][]{u_{#1}}
\newcommand{\uu}{\emox{\util}}

\newcommand{\uc}{\rendc{\uu}}
\newcommand{\ur}{\util[r]}

\newcommand{\urc}{\rendc{\ur}}

\newcommand{\vra}{\renda{\ur}}
\newcommand{\vrb}{\rendb{\ur}}
\newcommand{\vrc}{\rendc{\ur}}
\newcommand{\ui}{\util[i]}

\newcommand{\uic}{\rendc{\ui}}
\newcommand{\um}{\util[m]}

\newcommand{\umc}{\rendc{\um}}
\newcommand{\rr}{\emox{\rho}}
\newcommand{\ra}{\renda{\rr}}
\newcommand{\rb}{\rendb{\rr}}
\newcommand{\rc}{\rendc{\rr}}
\newcommand{\pp}{\emox{p}}
\newcommand{\pa}{\renda{\pp}}
\newcommand{\pb}{\rendb{\pp}}
\newcommand{\pc}{\rendc{\pp}}
\newcommand{\pii}{\emox{\pi}}
\newcommand{\pia}{\renda{\pii}}
\newcommand{\pib}{\rendb{\pii}}
\newcommand{\pic}{\rendc{\pii}}
\newcommand{\pdf}{\emox{\mu}}
\newcommand{\pda}{\renda{\pdf}}
\newcommand{\pdb}{\rendb{\pdf}}
\newcommand{\pdc}{\rendc{\pdf}}
\newcommand{\cdf}{\emox{\mathcal{M}}}

\newcommand{\cdb}{\rendb{\cdf}}

\begin{document}
\selectlanguage{british}
%
%
\title{Incentive Systems in Multi-Level Markets for Virtual Goods}
\author[A.\ U.\ Schmidt]{Andreas U.\ Schmidt} 
\thanks{\textit{Classifications. MSC} 91B60, 90B60, 46N10.  \textit{JEL} C51, C67, D4. 
\textit{ACM} K.4.4}
\thanks{\textit{Address:} Fraunhofer--Institute SIT, Dolivostraße 15,
  64293 Darmstadt, Germany} \thanks{\textit{E-Mail Address:}
  \href{mailto:Andreas.U.Schmidt@sit.fraunhofer.de}{Andreas.U.Schmidt@sit.fraunhofer.de}}
\thanks{\textit{URL:} \url{http://www.math.uni-frankfurt.de/~aschmidt}}
\begin{abstract}
  As an alternative to rigid DRM measures, ways of
  marketing virtual goods through multi-level or
  networked marketing have raised some interest.  This report is a
  first approach to multi-level markets for virtual goods from the
  viewpoint of theoretical economy.  A generic, kinematic model for the
  monetary flow in multi-level markets, which quantitatively describes
  the incentives that buyers receive through resales revenues, is
  devised.  Building on it, the competition of goods is examined in a
  dynamical, utility-theoretic model enabling, in particular, a
  treatment of the free-rider problem.  The most important
  implications for the design of multi-level market mechanisms for
  virtual goods, or \textit{multi-level incentive management
    systems}, are outlined.
\end{abstract}
\maketitle
%
%
\vspace*{-0.5cm}
\section{Introduction}\label{sec:ïntro}
\textit{Information goods} share the attributes of transferability and
non-rivalry with public goods, and additionally are durable\ie show no
wear out by usage or time~\cite{b:SV99,STEG04}.  Like with a private
good, however, original creation can be costly, whereas reproduction and 
redistribution are cheap.  This is the more true for
\textit{virtual goods}~\cite{AH03}\ie information goods in intangible,
digital form, which are distributed through electronic networks.
Free-rider phenomena and ``piracy'' plague their creators and
distributors, a problem which is conventionally approached using copy
protection measures and/or digital rights management (DRM) systems
which, generally speaking, aim at restoring some of the features of
private, physical goods.  This practise, backed by WIPO
treaties~\cite{WCT} and national copyright law in signatory states
giving DRM techniques protected legal status, has aroused public
controversy and an ongoing scientific discussion about its various
fundamental~\cite{DRM}, economic~\cite{KIN03}, and pragmatic
problems~\cite{STW04}, cf.~\cite{IEP-SI04} for a more general
discussion of the underlying concepts of intellectual property rights.
The general legitimacy of DRM measures which tend to disrupt
consumers' expectations on their individual usage of the
good~\cite{MHB03}, is doubtful in light of empirical findings on the
effect of illegal file-sharing on record sales~\cite{OS04}, which
seems negligible.
Therefore, as an alternative to the protection of virtual goods by
DRM, so called incentive management (IM) systems have recently
emerged.  They promise to yield a fair remuneration to the originator
of the good, who may be identical with its creator or not, without
necessitating copy protection or disruption of users' expectations on
``fair'' and ``personal'' uses.  One of the first such systems, and
one which is already in practical use is the so called Potato
System~\cite{GNACM02,GNWD03}.  It is based on super distribution of
the virtual good from buyer to buyer, whereby each buyer obtains,
along with the good itself, the right to redistribute it on
commission.  Upon resale, she will obtain a share of the purchase
price as an additional incentive.
The rationale behind this kind of scheme, called here
\textbf{multi-level IM systems}, is as obvious as appealing.  Rather
than to discourage illegal distribution of the good by more or less
unpopular measures, the aim is to make legal distribution more
attractive than ``piracy''~\cite{FET}.  Concurrently, the scheme
purports to attribute a fair remuneration to the party from which the
good originated, for instance the creator of a work of which the
virtual good is an embodiment.

The present report contributes a building block to the presently
lacking study of multi-level IM in the framework of theoretical
economy. Section~\ref{sec:flow_model} introduces a simple model for 
the monetary flux in a general multi-level market and derives the most
basic results pertaining to it. The model is complemented by a dynamical
model for the competition of two goods in such a market in 
Section~\ref{sec:competition_model}, with particular consideration of
virtual goods.
Two important special cases are treated in Section~\ref{sec:cases}.
Section~\ref{sec:fr} covers the free-rider problem and the 
potential of multi-level IM to counter it, while Section~\ref{sec:competition-sim}
gives a first account of genuine competition between two goods.
Section~\ref{sec:discussion} offers a qualitative discussion of the
issues raised in the preceding theoretical ones.
It is argued
in Section~\ref{sec:fair} that,
judged on grounds of the theoretical analysis,
multi-level IM can be a fair 
scheme despite its similarity to pyramid schemes.
The free-rider problem is recast as an issue of information economy
in Section~\ref{sec:free-rider}.
Section~\ref{sec:dyn_price} offers some thoughts on the general potential
of multi-level IM to influence markets through  determining 
the incentive via dynamical forward pricing.
A particular problem of multi-level markets, namely cannibalisation
by a potent reseller, is alluded to in Section~\ref{sec:root}.
Section~\ref{sec:conclusions} concludes by noting some directions
for further work. Proofs, and some technical material, are
contained in Appendices~\ref{sec:proofs}, \ref{sec:weibull-rho},
respectively. Figures can be found at the end of the paper. 
\section{Monetary Flux Model}\label{sec:flow_model}
The model we devise is continuous and kinematic in nature.  
That means firstly, that we
describe all pertinent quantities by variables with continuous range.
Secondly, that it describes the monetary flux between
the market players, and other relevant quantities, such as the
expected resales revenue, are to be derived from the kinematics.

About the market players no special assumptions are made, in particular
with regard to their decision making processes.  
That is, the model is neutral with respect to the
detailed structure of the monopolist firm marketing the good (which we
called its originator), and the consumer buying it.  Thus the agents
are solely discriminated by the time $t$ at which they
enter the market\ie
buy the good from another agent already present in it.
Consequently, buying the good happens only once for each agent, while
resale can happen to arbitrary amounts at subsequent times.  The
market in turn is assumed to be homogeneous\ie all agents have equal
probability to trade with each other.  In accordance, no
special market dynamics is assumed by letting the number $n(t)$ of agents in the
market at time $t$ be an unspecified function with continuous,
non-negative, finite or infinite range.  The
resales price at time $t$ is denoted by $\pi(t)$.

The expected (average) monetary incentive $v_i$ for an agent entering 
the market at time $t$ is given by
\[
v_i(t)=v_r(t)-\pi(t),
\]
that is, the expected revenue $v_r$ from resales to agents entering
the market at later times, diminished by the price at which the good
was bought\ie the resales price at time $t$.  To calculate $v_r$,
note that the influx of agents into the market is given by
$\dot{n}(t')=\dd n(t')/\dd t'$ at any later time $t'>t$, and if the
agent was alone then one could integrate $\pi(t')\dot{n}(t')$ over an
interval to obtain the resale revenue accumulated in it.  But since
there is competition in the reseller market, and all $n(t')$ agents
have equal probability to strike a deal with the newcomers, the
integrand must be divided by $n(t')$.  Thus
\[
v_r(t)=\int_t^\infty\frac{\pi(t')}{n(t')}\dot{n}(t')\,\dd t'.
\]
Reparametrisation by the
monotonously increasing number of agents $n(t)$,
makes the independence of the market dynamics manifest and yields
\[
v_r(n)=\int_n^{n_\infty}\frac{\pi(n')}{n'}\,\dd n',
\]
in which the market size $n_\infty$ may be finite or infinite.

The model neither specifies all the endogenous and
exogenous dynamically changing 
factors that may contribute to a complete model of
multi-level markets, nor does it presume any special estimators for them.
Accordingly, the fundamental price function $\pi$, as well as the
market dynamics, is left completely unspecified and can be generated
by any underlying mechanism without affecting any general results
derived from the model.

It is instructive to solve the homogeneous equation
$v_i=0$, corresponding to an expected balance between resales revenues
and buying price.  In this case, $\pi$ would necessarily
have to satisfy the differential equation
${\dd \pi(n)}/{\dd n} = -{\pi(n)}/{n}$,
the unique solution of which is $\pi(n)=\pi(0)/n$.  With this solution
however,  one obtains $v_i=-\pi(0)/n_\infty$, showing that this $\pi$ is not a
solution of the homogeneous equation for $n_\infty<\infty$.  The same
reasoning applies to any constant, nonzero $v_i$ and it follows that
such a situation is not realisable in a finite market,
due to the singular nature of the integral operator defining $v_i$.
Thus it makes sense to specialise to finite markets\ie to take $n_\infty<\infty$.  
Then, a nonsingular re-parametrisation can be applied, replacing $n$ with the
market saturation $s=n/n_\infty$, $0\leq s\leq1$.  
The integral operator $K\colon \pi \longmapsto v_i$, a Volterra operator of the
second kind, is defined by
\[
(K\pi)(s)\DEF v_i(s)=\int_s^1\frac{\pi(s')}{s'}\,\dd s'-\pi(s).
\]
As this operator describes a closed market, one would expect it to
satisfy a conservation law. In the present case this law takes the
form of a game-theoretical zero-sum condition.
\begin{Lprop}[\textbf{Zero-Sum Condition}]{zero-sum}
  If $\pi$ is bounded then
\[
\int_0^1 v_i(s)\,\dd s =0.
\]
\end{Lprop}
The proofs of all statements are easy calculations and are deferred to
Appendix~\ref{sec:proofs}.
The zero-sum condition expresses that wins and losses in incentive
compensate each other exactly.  It is also the main reason why the
attempt to obtain a nontrivial solution to the homogeneous equation
was bound to fail (notice that $\pi=\pi(0)/n$ is too singular at $0$ to
fall in the scope of \refP{zero-sum}).
One important feature of the model is that the incentive is \textit{scale-free}\ie
does not depend on $n_\infty$.

For regular enough $\pi$, the inverse of $K$ is easily obtained as a
solution of the inhomogeneous equation $K\pi=v_i$. The
derivatives of $\pi$, $v_i$, are denoted by $\dot{\pi}$, $\dot{v}_i$,
respectively.
\begin{Lprop}{inverse}
  $K$ maps $\mathcal{V}\DEF C^1([0,1])$ bijectively onto
\[
\mathcal{W}\DEF 
\left\{ v_i\in C^1((0,1]) \Bigm| {\textstyle\int\nolimits_0^1}v_i=0,\ 
  v_i=o({\textstyle\frac{1}{s}}),\text{ and }
  \dot{v}_i=O({\textstyle\frac{1}{s}})\ (s\to0) \right\}.
\]
The inverse of $K\colon\mathcal{V}\longrightarrow\mathcal{W}$ is
\begin{equation}
  \label{eq:inverse}
(\check{K}v_i)(s)\DEF-\frac{1}{s}\int_0^s \sigma\dot{v}_i(\sigma)\,\dd\sigma.  
\end{equation}
\end{Lprop}

Although nothing in principle prevents a forward monetary flow from
earlier market entrants to later ones by negative prices $\pi<0$, 
the more conventional case is that of
positive resale prices.  According to the inversion formula in
\refP{inverse}, it is sufficient that $\dot{v}$ is non-positive for
$\pi$ to remain non-negative, that is positive (non-negative) prices
are always obtained if the incentive is (strictly) monotonic
decreasing.  
The necessary and sufficient condition for positive prices
reads as follows.
\begin{Lprop}{sharp-positive}
  Let $\pi\in C^1([0,1])$.  Then, $\pi$ is positive if and only if
\[
\frac{1}{s}\int_0^sv_i(\sigma)\,\dd\sigma>v_i(s) \quad\text{for all }s.
\]
\end{Lprop}
This result has a rather direct interpretation.  It says that the
monetary flow is always directed backwards if and only if the expected
incentive at a certain time is smaller than the average expected
incentive before that time.

The basic model can easily amended by further features.
In particular it is desirable to take transaction costs and a
commission into account.
The former can be easily incorporated as follows.
In the resale process, the buyer as well as the seller can incur
transaction costs.  We show how both of these additional costs
can be incorporated in the model when they are constant.
While
the buyer's transaction cost $\beta\geq0$ directly adds to the price
$\pi(s)$ and can therefore be absorbed in that function, the seller's
transaction cost $\sigma\geq0$ modifies the integrand for the calculation of
$v_r$ from $\pi(s)/s$ to $(\pi(s)-\sigma)/s$.  Upon integration, this yields
a negative contribution in the incentive of the form
\[
v_i(s)=\int_s^1\frac{\pi(s')}{s'}\,\dd s'+\sigma \ln s -\pi(s).
\]

If there is an entity, called the \textbf{collector},
which collects part of the resales revenue\eg
to remunerate the creator of the good, and pays only part of it as a 
commission to resellers, the market turns into an open system.
The \textbf{commission factor} $0\leq\gamma\leq1$
diminishes the revenue of a single resale from $\pi$ to $\gamma\pi$,
and the modified operator $K_\gamma$ yielding the incentive
$v_{i,\gamma}$ becomes
\[
(K_\gamma\pi)(s)=\int_s^1\frac{\gamma(s')\pi(s')}{s'}\,\dd s'-\pi(s).
\]
Its inverse for differentiable $\pi$ can still be calculated as
\begin{equation}
  \label{eq:commission_inverse}
(\check{K}_\gamma v_{i,\gamma}) (s)=-\ee^{-\int_0^s\frac{\gamma(\tau)}{\tau}\,\dd\tau}
\int_0^s\dot{v}_{i,\gamma}(\sigma)\ee^{\int_0^\sigma\frac{\gamma(\tau)}{\tau}\,\dd\tau}\,\dd\sigma.  
\end{equation}

For constant commission this reduces to
\[
(\check{K}_{\gamma=\text{const.}} v_{i,\gamma}) (s)=-\frac{1}{s^\gamma}
\int_0^s\dot{v}_{i,\gamma}(\sigma)\sigma^\gamma\,\dd\sigma.
\]
The amount of money $v_{c,\gamma}$ taken out of the market by the
collector can be calculated\eg when $\pi$ is bounded, as in
\refP{zero-sum} to obtain
\[
v_{c,\gamma}=\int_0^1 \bigl((K-K_\gamma)\pi\bigr)(s)\,\dd s=
\int_0^1\bigl(1-\gamma(s')\bigr)\pi(s')\,\dd s',
\]  
as expected.  Note that this quantity is still normalised and the
gross commission collected is $n_\infty v_{c,\gamma}$.  The market with
commission no longer satisfies the zero-sum condition but rather its
analogue
\[
\int_0^1v_{i,\gamma}(s)\,\dd s=-v_{c,\gamma},
\]
balanced with the collector's share.

\begin{figure}[p]
  \centering 
  \resizebox{0.8\textwidth}{!}{\includegraphics{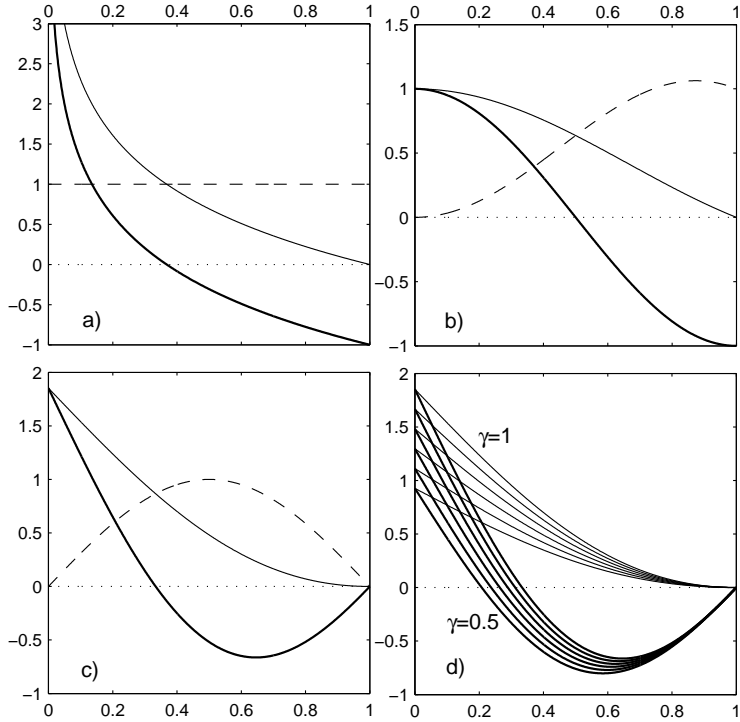}} 
  \caption{%
    Examples for prices $\pi$ (dashed), expected resales revenues $v_r$
    (thin solid), and incentives $v_i$ (thick solid).  a) $\pi=1$,
    $v_r=-\ln s$, $v_i=-\ln s -1$.  b) $\pi(s)=\sin(\pi s)/(\pi
    s)-\cos(\pi s)$, $v_r=\sin(\pi s)/(\pi s)$, $v_i=\cos(\pi s)$.  c)
    $\pi(s)=\sin(\pi s)$, $v_r=\Si(\pi)-\Si(\pi s)$, $v_i=\Si(\pi)-\Si(\pi
    s)-\sin(\pi s)$, where $\Si$ is the integral sine function.  d)
    Price as in c) with commission factor $\gamma$ varying from $1$ to
    $0.5$ in steps of size $0.1$.}
  \label{fig:incentive_examples}
\end{figure}

A continuous model is an idealisation of a realistic
market where buyers enter one by one\ie the market size evolves in
discrete steps.  This entails artifacts, 
most notably the logarithmic singularity for $v_i(s)$ as
$s\searrow0$ when $\pi(0)>0$, see Figure~\ref{fig:incentive_examples}~a). 
Therefore one needs to examine the
discrepancy between the incentive obtained from the continuous model
and the one calculated by discrete summation somewhat more closely.
For a constant price $\pi(s)=\pi$, the discrete model can be solved
directly.  Agents are labelled with $k=1,\ldots,n_\infty$, by the order of
market entrance, and this yields for the expected incentive
$\overline{v}_i$ of the discrete case
\[
\overline{v}_i=\pi\left(\sum_{k'=k+1}^{n_\infty}\frac{1}{k'-1}-1\right)=
\pi\left(\Psi(n_\infty)-\Psi(k)-1\right),
\]
where the Digamma function $\Psi(z)=\Gamma'(z)/\Gamma(z)$ is the logarithmic
derivative of the Gamma function, see~\cite[p.~39]{OLV74}.

In the general case, we have to look at the difference between
$v_i(s)$ and the discrete incentive $\overline{v}_i(s\cdot n_\infty)$ at the
corresponding point.
\begin{Lprop}{asymptoticbound}
  For bounded, non-negative $\pi$ holds
\[
\ABS{v_i(s)-\overline{v}_i(sn_\infty)}\leq
\frac{\pi_\mathrm{max}}{2}\left[%
  \frac{1+s}{sn_\infty}+ \frac{1}{6}\frac{1+s^2}{(sn_\infty)^2}+
  O\left(\frac{1+s^4}{(sn_\infty)^4}\right) \right],
\]
with $\pi_{\mathrm{max}}\DEF\max_{s\in[0,1]}\pi(s)$, and in which the
term of order $(sn_\infty)^4$ is strictly dominated by the previous one.
\end{Lprop}
Te error behaviour of the continuous model is
rather benign in that it decays with the inverse of the market size at
any finite saturation $s>0$.  For fixed $k=sn_\infty$ on the other hand, a
constant error bounded by $c_k\pi_{\mathrm{max}}$ for some $c_k>0$,
will always remain.

The dynamics of multi-level markets are prone to be influenced, if not dominated,
by network effects~\cite{ECON96}, and it is desirable to assess how the incentive 
relates to them.
Network effects are
understood in the economics literature as the benefit that accrues to
a user of a good or a service because he or she is one of the many who
use it.  Simple functional forms of network effects for special types
of networks\eg telecommunication networks, 
such as Sarnoff's, Metcalfe's, and Reed's law, are often
taken as heuristics to explain the dynamics of the growth of networks
of the respective type.  The most prominent phenomena traced back in
this way to network effects are a ``slow startup'', the
existence of a ``critical mass''~\cite{LCP03}, and strong
(supra-exponential) growth after this mass has been reached.  Models
for network externalities and their effects on prices and utility are
numerous and detailed, see\eg~\cite{ECON96a,SUN03} and references
therein, while global models, such as~\cite{SWA02} for the possible
functional forms of network externalities, are scarce.

Network utility can spatially either be understood in a global sense
as the \textbf{aggregate} value, summed over all members of the
network, or as the local, \textbf{individual} value enjoyed by its
single members.  In the present context, each case is in turn subdivided on
the temporal axis into the \textbf{dynamic} utility given as a
function of the saturation $s$, as a relative variable, and the
\textbf{kinematic} utility, which is the scaling behaviour of the
utility with the market size $n_\infty$.

The aggregate utilities are the simpler ones to discuss.  
In fact, the only kinematic
aggregate utility in our model is that
obtained by the replication of the good and distribution of it
to the members of the network, a contribution which is always of order
$O(n_\infty)$, like in broadcast networks.  The incentive contributes to
aggregate utilities only in a dynamic way, since it is given by
\[
n_\infty\cdot\int_0^sv_i(s')\,\dd s',
\]
which approaches zero for $s\to 1$ due to the
zero-sum condition, or is of the order $O(-n_\infty)$, more
precisely $-n_\infty v_{c,\gamma}$, if a commission is in effect.

The dynamic, individual utility of the network is
directly affected by $v_i$.  
In fact, in the continuous model there is no other relevant external contribution 
to individual utility, since the kinematic, individual utility, defined
as the scaling behaviour of $v_i$ with $n_\infty$, is $O(1)$ precisely if
$\pi$ is $O(1)$ $(n_\infty\to\infty)$\ie if the price stays bounded as the
market grows.  While this argument holds for large saturations,
some care has to be taken for low ones.  
Firstly, it might be that the continuous model
introduces artifacts that lead to nontrivial scaling for small
$s=k/n_\infty$, keeping $k$ fixed while letting $n_\infty$ grow.  This is
however excluded by the error bound obtained in \refP{asymptoticbound}.
The scale-free behaviour of the continuous model is therefore stable
for nonzero $s$.  
For small, fixed $k=sn_\infty$, and if $\pi(0)>0$ a scaling of the kinematic, 
individual utility of order $O(\ln n_\infty)$ is obtained.  This
is in accordance with the conventional wisdom that in pyramid schemes
the profiteers realise profits which scale logarithmically with the
number of participants. 
In conclusion, the incentive introduces a
single, independent network externality which,
except for a rather moderate effect on early
subscribers, does not exert a strong effect on the market.
This was to be expected since the market
describes has no special structural properties.

Figure~\ref{fig:incentive_examples} a) shows the most basic example
of resales revenues and incentives resulting from a constant price.
It exhibits the logarithmic singularity present in the continuous model,
and which will always emerge if $\pi(0)$ is positive.
The singularity is avoided if $\pi(0)=0$ as in b) and c).
Additionally, in c) the incentive is forced to zero as $s\to 1$ by letting
$\pi$ approach zero, and also
shows a case where $v_i$ is not always monotonic decreasing and $\pi$
is still positive.
The effect of a commission factor is exhibited in 
Figure~\ref{fig:incentive_examples} d).

\section{Competition Model}\label{sec:competition_model}
To devise a dynamical model for the competition of two goods, say $A$ and $B$, in a 
multi-level market described by the model above, 
an utility-theoretic approach is suitable.
Let \scc ($\bullet=A$ or $B$) denote the partial market sizes, or
\textbf{market shares} for good
$A$, and $B$, respectively. As all other variables introduced
below, they are considered as dependent
variables $\scc=\scc(s)$ satisfying $\saa+\sbb=s$.
This account manifestly treats $\denota$ and $\denotb$ as
substitute goods\ie agents decide exclusively for either one 
or the other.

To describe the decision probability $\rc=\rc(s)$ for buying $A$ or
$B$, respectively, at saturation $s$, at least three factors need to be
taken into account.
The first is the distribution of the genuine, individual utilities 
$\uc$ of the good across the population. 
The second is the individual utility $\uic\DEF\urc-\pic$
originating from individual utilities $\urc$ arising from 
expected resales revenues, where $\pic=\pic(s)$
is the price of the respective goods.
In the present model these two factors are considered as exogenous ones,
while the third one is an endogenous, generic network effect, captured
in a contribution $\umc$ to the utility.
It is convenient to introduce, for all utilities,
the \textbf{bias} $\Delta x \DEF x^{\denota}-x^{\denotb}$ as a measure for the advantage
gained by deciding for $\denota$ rather than $\denotb$.   

Let $\pdc=\pdc(\uc)$ be the probability density function (PDF) of the
distribution of \uc across the population.
The distributions for
both goods are taken to be equal and to depend only on the
respective \textbf{popularities} $\pc\geq0$\ie
$\pdc(\uc)=\pdf(\pc,\uc)$.
We assume that $\pdf(x)=0$ for $x<0$, and that \pdf satisfies the
principle of stochastic dominance\ie
\[
\cdf(q,x)\geq\cdf(p,x)\quad\text{for }p\geq q,
\]
where $\cdf(\pp,x)=\int_0^x \pdf(\pp,y)\dd y$ is the cumulative density function
(CDF) of \pdf.
With these settings, the probability that an agent decides to
buy $\denota$ is $\ra(\Delta)\DEF\Pr(\Delta u + \Delta > 0)$, where
the \textbf{decision bias} $\Delta$ subsumes all other utility contributions to
the bias for $\denota$.   
It follows, with the notation $\ra(\pa,\pb;\Delta)=\ra(\Delta)$, making the
dependency of \ra on the popularities explicit,
\begin{equation}
\label{eq:rhodef}
  \begin{aligned}
  \ra(\pa,\pb;\Delta) &= \int_0^\infty \dd \pda(u) \int_0^{u+\Delta} \pdb(u') \\
& = \int_0^\infty \dd \pda(u) \cdb(u+\Delta) =  \int_0^\infty \dd \pda(u) \cdf(u+\Delta-\pb).
\end{aligned}    
  \end{equation}
In simple models as used below, 
the distributions \pdc are given 
in translation form $\pdf(\pc,\uc)=\pdf(0,\uc-\pc)$, in which case 
\eqref{eq:rhodef} simplifies to
\begin{equation}
\label{eq:rhocalc}
\ra(\pa,\pb;\Delta) =     \int_0^\infty \dd \pdf(u) \cdf(u+\Delta+\Delta\pp),
\end{equation}
where $\Delta\pp=\pa-\pb$ is the \textbf{popularity bias}.
\begin{prop}
  \label{prop:rho}
The function $\ra(p,q;\Delta)$ is monotonously increasing in $p$ and 
$\Delta$, monotonously decreasing in $q$, and satisfies

\noindent i) $\ra(p,q;\Delta)=1-\rb(p,q;\Delta)$,

\noindent ii) $\ra(p,q;\Delta)=1-\ra(q,p;-\Delta)$,

\noindent iii) $\ra(p,q,\Delta)=\rb(q,p,-\Delta)$.
\end{prop}
Having at hand the probability 
$\ra(\sat)=\ra(\pa(s),\pb(s);\Delta(s))$
to buy $\denota$ at a given total saturation, 
we can write down the fundamental
relation governing the dynamics of the multi-level market in which $\denota$
and $\denotb$ compete.
\begin{equation}
  \label{eq:fundrel}
  \saa(s)=\int_0^s \ra(s')\dd s'.
\end{equation}

The second element contributing to the decision bias is the agents'
\textit{ex ante} estimation of resales revenues and the incentive, 
thus defining \urc and in turn the \textbf{resales revenue} and
\textbf{incentive bias} $\Delta\ur$ and $\Delta\ui=\Delta\ur-\Delta\pii$, respectively. 
Due to limited knowledge about the market situation, agents are bound to
behave according to a rule of bounded rationality and using partial information. 
We choose $\urc(s)\DEF\vrc(s)\cdot\rc(s)$, where $\vrc$ is the bare resales
revenue $\vrc(s)=\int_s^1 \pic/s'\dd s'$. 
Here, $\ra(s)=\ra(\pa(s),\pb(s);0)$, and $\rb(s)=\ra(\pb(s),\pa(s);0)$
are the probabilities for buying $\denota$, $\denotb$, respectively,
governed merely by popularity.
That is, agents expect to gain the resales revenue of an undisturbed
multi-level market of relative size $\rc(s)$.
Sellers transaction costs, which can be assumed to be of 
similar magnitude for both goods, and small for virtual ones are 
neglected, as well as commissions by which we focus on the competition
between the goods, exclusively.
The assumptions on the agents' accessible information
underlying this Ansatz are i) the price schedules $\pic(s)$ 
are public knowledge, ii) $s$ can be estimated with good precision, 
as well as iii) $\rc(s)$.
While i) depends on the mechanism implemented by the multi-level IM system,
ii) and iii) can be justified to the end that they represent information
accessible through \textit{local measurements} within an agent's communication
reach.
Summarising, this definition of \urc represents partially but 
rather well informed individuals which behave subjectively rational.
Further discussion of \urc is contained in Sections~\ref{sec:fair}
and~\ref{sec:free-rider}.

As already alluded to in Section~\ref{sec:flow_model}, the dynamics
of multi-level markets is very likely to be affected by network effects.
In fact, in a completely homogeneous  market and 
in the absence of other externalities influencing an agent's
decision, a network effect becomes dominant.
For, if resellers of good $\denota$, say, are rare then  a
buyer will be very likely to buy from a reseller of $\denotb$.
In such a situation \ra can become negligible and the market completely
governed by the \textbf{multiplier effect} of resellers of $\denotb$.
We do not presume such an extreme effect to be prevalent, and, since
generic utility-theoretic treatments of network effects are lacking
except for special cases, cf.~\cite{ECON96a,LCP03,SUN03}, we choose an
\textit{ad hoc}, moderate multiplier utility 
$\umc\DEF\EPS\scc/\sat$ depending on an adjustable parameter $\EPS$. 
This yields a \textbf{multiplier bias} $\Delta\um=\EPS(\saa-\sbb)/ \sat=
\EPS(2\saa/ \sat - 1)$ as the single endogenous contribution to \ra.

With the specification
\begin{equation}
  \label{eq:deltaspec}
\Delta\DEF \Delta\ui+\Delta\um=\vra\ra-\vrb\rb-(\pia-\pib)+
\EPS\left(\frac{2\saa}{\sat}-1\right)  
\end{equation}
the model for the competition of two goods in a
multi-level IM market is complete.  
Note that~\eqref{eq:rhodef}, \eqref{eq:fundrel}, \eqref{eq:deltaspec} 
present an exactly solvable integral equation for \saa.
Will will now examine some special numerical solutions of it.
\section{Two Special Cases}\label{sec:cases}
Though the presented competition model is simple, the space of situations
covered by it is vast, as
input data are the price schedules \pic, popularity functions \pc, and the
multiplier factor coupling $\EPS$, but also the dependency of \pdf on the
popularities.
Here we assume that the latter be of translation form~\eqref{eq:rhocalc},
and specify that $\pdf(0,u)$ is given by the special Weibull distribution
$f(u;1,2)$, see Appendix~\ref{sec:weibull-rho}, 
in which case \ra takes the analytical form~\eqref{eq:weibull-rho}.
For \pic and \pc we specialise to spike functions
\[
g(s;m)\DEF
\begin{cases}
  s/m, & \text{for $0\leq s\leq m$;}\\
  (1-s)/(1-m) & \text{for $m<s \leq 1$.}
\end{cases}
\]
Pragmatically, \pic of spike form
offer an early-subscriber discount and a late-adopter
rebate, cf.\
Section~\ref{sec:dyn_price}. 
Technically, they  are the simplest 
price schedules which avoid an initial singularity, thereby minimising 
the variance with a discrete model, and correspond to markets
closing at finite size.

Besides the market shares $\scc$ and the \textbf{final shares} 
$\rendc{S}\DEF\scc(1)$,
the \textbf{turnovers}
\[
\rendc{t}(s)\DEF 
\int_0^s\pic(s')\rc(s')\dd s'=
\int_0^s\pic(s')\rendc{\dot{s}}(s')\dd s'=
\int_0^{\scc(s)} \pic(\scc')\dd \scc' 
\]
and the \textbf{total turnovers} $\rendc{T}\DEF\rendc{t}(1)$ are important indicators
for the economic performance of the competing goods.
Note that the maximal turnover that a good can generate is $1/2$
for spike functions.
Furthermore, we examine the discrepancy between agents' expectation and the
actual resales revenue they can achieve, similarly calculated as
\[
\rendc{v_r}(s)\DEF\int_s^1\frac{\pic(s')}{\scc}\rc(s')\dd s'=
\int_{\scc(s)}^{\rendc{S}} \frac{\pic(\scc')}{\scc'}\dd  \scc',
\]
and the resulting actual incentive $\rendc{v_i}(s)\DEF\rendc{v_r}(s)-\pic(s)$.
\subsection{Free-Rider Phenomena}
\label{sec:fr}
To counter free-rider phenomena is the main aim behind the conception
of multi-level marketing of virtual goods.
In fact, the content distribution network of multi-level IM systems
like the Potato system~\cite{GNACM02,GNWD03,PS} is very similar to the
peer-to-peer networks commonly used by free riders~\cite{ZK04}.
By this rationale, we can compare the performance of a virtual good
$\denota$ with a pirated version $\denotb$ of it in the \textit{same}
multi-level market.
That is, the popularities are equal $\pa=\pb$ and $\denotb$ is free\ie
$\pib=0$.
Since in this case no confusion can arise, we sometimes drop the superscript
$\denota$.

\begin{figure}[p]
\centering
\includegraphics{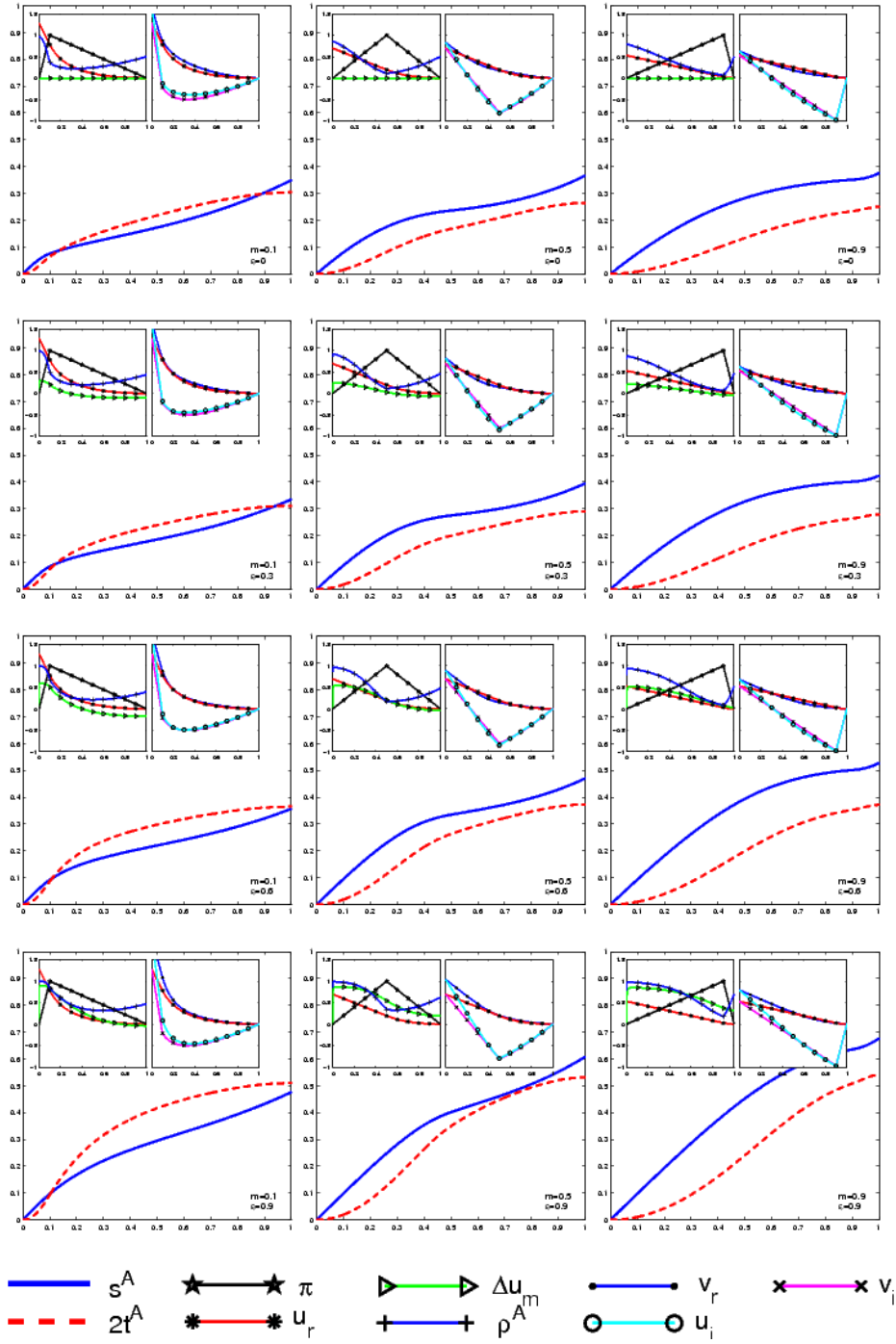}

\caption{Example market evolutions in the free-rider setting.}
  \label{fig:fr}
\end{figure}

Figure~\ref{fig:fr} shows how the market evolves in this setting for some
selected cases. 
The main figures show the market indicators $s$ and $2t$ (relative to 
the maximal turnover $1/2$).
The left and right inlays exhibit the factors contributing to the
decision biases, and the resulting $\rr$, respectively, 
a comparison between expected and actual resales revenues and
incentives.
The left column has an early peaking price schedule $m=0.1$,
entailing an initially very high and then steeply dropping incentive bias
(right inlays),
while in the centre and right columns $m=0.5$, $0.9$, respectively, 
which in turns leads to a smaller, but longer lasting positive initial
$\Delta\ui$.
Note the sharp negative peak of $\Delta\ui$ for $m=0.9$, 
entailing a significant entry deterrence\ie  $\rr<0.5$ at late
times.
The right inlays show that the simple rule for $\ur$ leads to good
estimations for $v_r$, and in turn $v_i$. Agents tend to underestimate
the resales revenues they can achieve at early times and overestimate
them only in an intermediate phase.
The increasing influence of multiplier effects can be observed 
along the four rows for which $\EPS=0$, $0.3$, $0.6$, $0.9$ from
top to bottom.

\begin{figure}[p]
  \centering\resizebox{\textwidth}{!}{\rotatebox{0}{\includegraphics{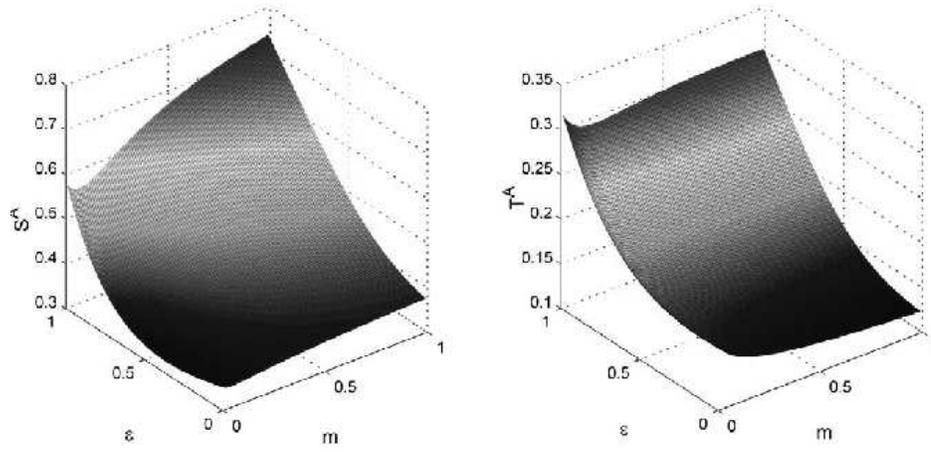}}}
  \caption{%
    Final shares (left) and total turnovers (right)
    in the free-rider setting.}
  \label{fig:frst}
\end{figure}

Even without a multiplier effect present, the incentive can lead to a non-negligible market share though not dominance.
However significant turnovers are not generated without exploiting
the multiplier effect by an initial invitation to enter\ie a positive
incentive at early times.
For multiplier biases $\EPS\cong 1$ which are comparable to the price and other
biases, good $\denota$ can reach market dominance and generate over $1/2$
of the maximum turnover.
To maximise turnovers, the price schedule must be aligned with the
market growth \saa, which is generally difficult.
Figure~\ref{fig:frst} shows the plateaus of $\renda{S}$ and $\renda{T}$ 
in dependence of $m$ and $\EPS$.
It can be seen that maximisation of turnover and share are conflicting goals.
\subsection{Smash Hits and Sleepers}
\label{sec:competition-sim}
\begin{figure}[p]
\centering
\includegraphics{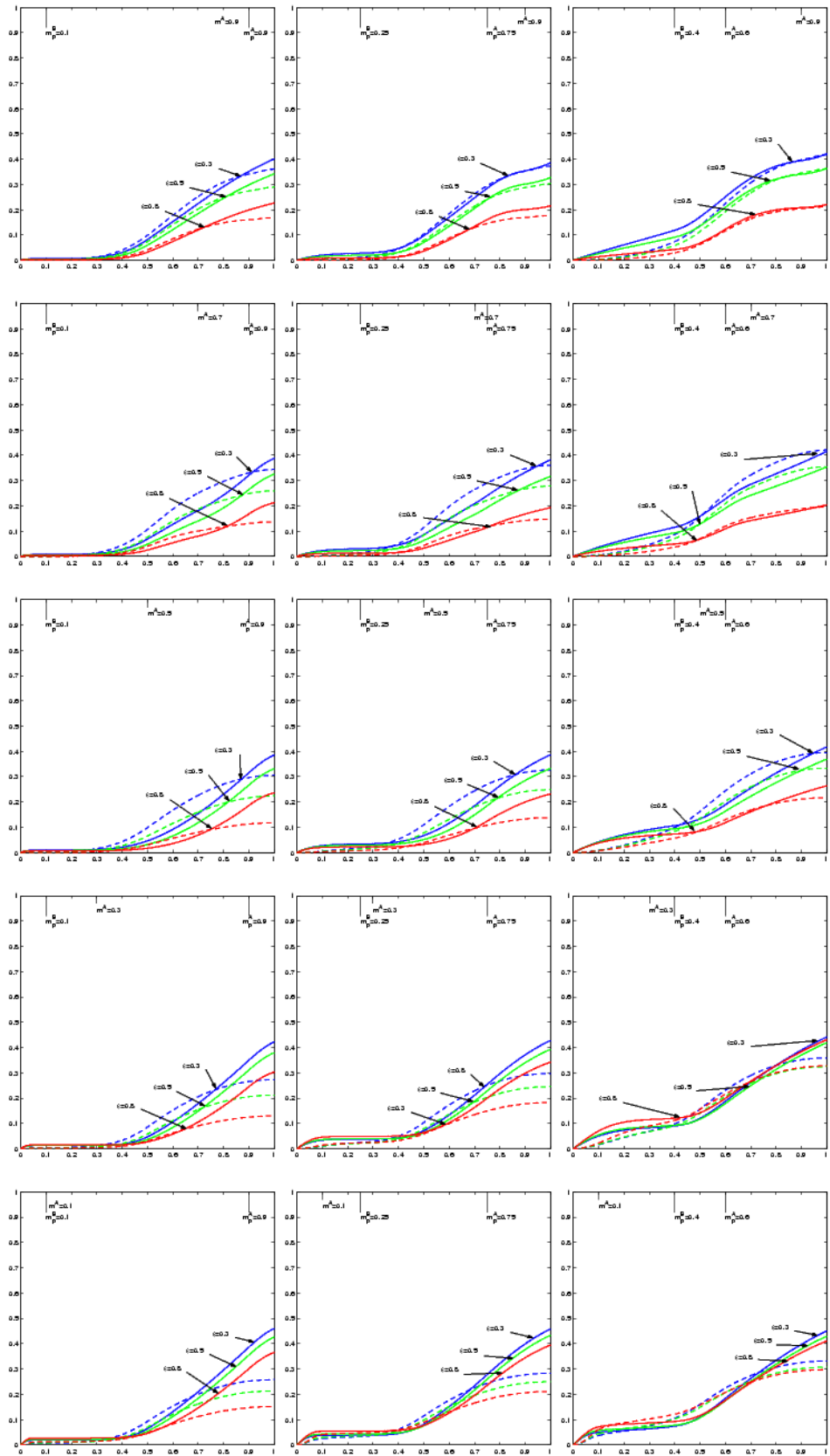}
\caption{Competition between sleeper and smash hit. Market shares $\saa$ 
are solid, turnovers $2\renda{t}$ dashed.}
  \label{fig:c}
\end{figure}

Scenarios for the competition of two goods are manifold within our model
and lack of space prohibits a comprehensive treatment.
Here, as a familiar example, good $\denota$ is assumed to
have a popularity function peaking later than that of $\denotb$\ie
$\denota$ would commonly be termed a `sleeper' while $\denotb$ can be
considered a `smash hit'. Denote by $\renda{m_p}$, $\rendb{m_p}$ the
popularity peaks of $\denota$ and $\denotb$, respectively.
The originator of $\denota$ would like to counter the slow startup
effect due to later popularity utilising an appropriate price schedule,
corresponding to various positionings of the peak $\renda{m}$ of his price
function. The price function of $\denotb$ is assumed to be centred,
$\rendb{m}=0.5$.
Examples are shown in Figure~\ref{fig:c}.
It can be seen that the final share of $\denota$ is mostly small if the
multiplier effect is strong, since then the early rise in popularity
of $\denotb$ gives $\denotb$ a persistent advantage.
To counter this by a long lasting rebate\ie a late price peak $\renda{m}$
is in fact possible, as the first two rows ($\renda{m}=0.9$, $\renda{m}=0.7$,
respectively) show.
The opposite strategy to start the market by an early peaking price and
therefore high initial incentive can also work, as can be observed in the
last two rows  ($\renda{m}=0.3$, $\renda{m}=0.1$,
respectively).
However in this case, the price function of $\denota$ is misaligned with the
market evolution and hampers the generation of turnovers.
In conclusion, to optimise the price function of the sleeper so as to
obtain good market shares \textit{and} turnovers, is difficult.
\section{Discussion and Practical Implications}\label{sec:discussion}
\subsection{Pyramid Schemes and the Issue of Fairness}
\label{sec:fair}
Attractive as it may be, multi-level IM has, some similarity with
pyramid sales schemes --- a publicly discouraged enterprise, which is
illegal under most jurisdictions.  Thus the natural question emerges,
whether IM systems based on super distribution on commission are a
tenable market mechanism at all, and in particular for virtual goods.
In practise, the question is whether multi-level IM falls in the
economical category of legitimate multi-level marketing (also referred
to as direct, or network sales), or of illegal pyramid
schemes~\cite{FTC}.

The key argument for the defence of multi-level IM is that a buyer acquires not
only a void right to resale, but with it a good of positive value,
meaning that potential losses he will incur when he enters the market
too late\ie too close to saturation to obtain significant resales
revenues, can be partially alleviated by the good's value.  

An important difference between pyramid schemes operating with
physical goods and multi-level IM is clarified through the analysis of
transaction costs.  The negative contribution of
sellers' transaction costs $-\sigma\ln s$ can hit early buyers very hard,
since they have to process many resales.  A particular case in the
real world to which this finding corresponds is the detrimental effect
of \textit{inventory loading} in pyramid schemes.  There~\cite{LL}, resellers
of the good incur extraordinarily high transaction costs by being
required too keep a large, non-returnable stock of the goods, probably more than they
could ever expect to sell.  The penalty  arising from this 
multiplies for early adopters
who actually sell a portion of the goods and are required to reorder
stock, which is then usually possible only in overly large lots.
Virtual goods are much tamer in this respect, since
 the marginal cost for their
replication, as well as the transaction cost for resales, are
small, if not negligible.  
Stock keeping in itself
is not an issue, since virtual goods allow for
principally infinite, lossless replication.  Marginal costs for their
replication and redistribution are, more often than not, orders of
magnitude smaller than their value, even if they are embodied on a
physical media like a CD, say, for transport.  This is the key
difference which makes multi-level IM of virtual goods more viable and
acceptable in many cases than analogous multi-level marketing schemes
for physical goods. 
 In the Potato system for instance, the processing
of resales, including accounting, billing, and charging is fully borne
by the central server, for which a percentage of the price is assigned
to the system~\cite{PS}.  That is, the transaction costs are absorbed
in the commission factor and after a buyer has received his resale
link from the system in a one-off transaction, the marginal costs for
resales are close to zero.

Thus, the individual utility of the good for the
buyers is central to the question of fairness of multi-level IM. 
If the good's utility is close to zero, then the scheme
cannot be considered fair anymore but resembles a pure Ponzi scheme or
``Peter-and-Paul'' scam.
Formalising, to be fair a multi-level IM scheme should meet the requirement
$\ui\geq-u$ (on average over the buyer population) if fairness is judged on a
subjective level, or $v_i\geq -u$ judged from a forensic perspective.
This condition limits the scope in which the incentive can be predetermined
using dynamical forward pricing in multi-level IM, see Section~\ref{sec:dyn_price}
below.
\subsection{The Free-Rider Problem}
\label{sec:free-rider}
A secondary meaning conventionally attributed to the term incentive,
is that the incentive can be used by the principal who places it as a
means to eventually meet some ends, in particular to minimise the
readiness of agents to take moral hazards, for instance becoming an
illegal free-rider~\cite{LM}.

Whether multi-level IM can be successful in meeting 
the aim to fully replace copy protection measures and
conventional DRM is a question for theoretical economy, cf.~\cite{DY04} 
for a treatment of this question in conventional market settings.
If the good is freely available, as, for instance, in the Potato system,
then it is not \textit{a priori}
clear that another equilibrium apart from $\renda{S}=0$ (only free riders)
exists.
The zero-sum condition tells us
that, globally, an agent partaking in the IM market is not worse off
on average than one not doing so, and thus a market of any size
$n_\infty>0$ is in fact a global 
equilibrium.  
Assume the
agents to be rational and conservative in the sense that they would
tend to copy the good for free in the
absence of an additional payoff.  
Then, a necessary condition for the
market to evolve is that there is an initial phase of in which
they can expect a positive pecuniary incentive, that is $\ui(s)>0$ for
$s<s_0$, $s_0>0$.

It is here that the free-rider phenomenon is closely connected
to the issue of fairness and the economical purport of information.
For if the zero-sum condition is \textit{common 
knowledge}, then rational agents would always choose
the free good since they know that later potential buyers with
negative incentive (actual or subjective) will do so.
This renders the success of real pyramid schemes paradoxical,
and shows that the incentive schedule is at most \textit{public
knowledge}: There must be agents who know that some others will have
a negative incentive but expect them to enter the market nonetheless.
This is the reason for modelling the decision mechanism of agents 
using a rule of bounded rationality, as in Section~\ref{sec:competition_model}.
As shown in Section~\ref{sec:fr}, an initial invitation to enter through a
positive incentive can, within the scope of this model and if 
combined with a (small) multiplier effect,
turn multi-level IM into a functioning tool to counter
the free-rider phenomenon.

The presence of the free version 
can be seen as an exogenous factor negatively affecting
the individual utility $u$, 
and in turn the scope for the determination of the incentive.
This is the classical dilemma for the marketing of digital goods and
the one addressed by copy protection and DRM.  To offer a pragmatical
conjecture, it might make sense to differentiate the freely
available version of the good from the one distributed through the
multi-level IM system, by adding some value \textit{and} copy
protection to the latter, though this would be a partial return to
conventional DRM measures, maybe in the ``softer'' forms of
watermarking, personalisation, and fingerprinting, to enable
traceability of illegal copies~\cite{DRM}.
\subsection{Dynamical Forward Pricing}
\label{sec:dyn_price}
For the operator of a multi-level IM system,
the primary goal to maximise the revenues for the originator of the
good through his share of the commissions, contains the secondary
sub-goal of promoting the distribution of the good\ie maximising the
market penetration.  A central, new result of the present study is the
possibility, via the inversion 
formulae~\eqref{eq:inverse} and~\eqref{eq:commission_inverse}, to
dynamically adapt the incentive during the evolution of the market if
the operator of the system controls the price as an external parameter.
This is useful to turn \textbf{multi-level IM systems with dynamical
  forward pricing} into tools for market mechanism design, to
achieve the mentioned goals.  While dynamical forward pricing is not
a new concept in the theory of information goods~\cite{JA04}, this
possibility has, in the context multi-level incentive markets for
virtual or physical goods, not yet been widely considered in the
literature.

Figure~\ref{fig:incentive_examples} shows the most basic possibilities
for price functions.
A constant price as in a) is associated with a strong favouritism of early buyers,
while later market entrants are increasingly penalised.
A typical example for what is conventionally termed an early
subscriber discount is shown in b).  
In real markets this is often used as a means to
spur the distribution of the good in an early stage of market
development\ie to counteract a slow startup effect.
For the marketing of virtual goods, such  an initial invitation to enter
becomes important, in
particular if the good is freely available through legal or illegal channels, 
and therefore early buyers cannot be sure about
their potential resales revenues which depend logarithmically on
the market size (remember that $v_i(k=sn_\infty)$ scales as $\ln n_\infty$).
The price associated with
the incentive  in b) is monotonous increasing, resulting in a double penalty to
later buyers who pay more and receive less incentive.
Buyers who enter this market for some reason at late times will notice
that they are disfavoured, and
possibly tend to become frustrated. 
The third example in Figure~\ref{fig:incentive_examples} c) improves on b)
by letting the price vanish when the market reaches saturation.
  This $v_i$ combines a discount for early
subscribers with a rebate for very late ones who finally obtain the
good gratuitously.  
This pricing can therefore potentially be used to
spur the distribution of the good in early market phases through low
price and high incentive, as well as at late times, when the good
itself might have lost in utility and the market looses dynamics.  If we
assume that the market has a positive growth dynamics in an
intermediate phase associated with a high demand and maybe a higher
individual utility, then it is also reasonable to let the prices peak and
lower the incentive during this phase, as in c). 
Deepness and position of the minimum of $v_i$ can be adjusted almost
arbitrarily. 
Finally, d) shows the relatively limited effect that a collector has on the
incentive. In particular it can be seen that the point at which the incentive 
becomes negative is not significantly shifted by the increasing commission factor.

For an implementation of dynamical forward pricing,
information on the market dynamics becomes essential, in particular the current
size $n(t)$ of the market must be known.  This is in fact the case\eg
in the Potato system, where a central server
counts every single acquisition of the good.  
A much more difficult to determine variable is the absolute market
size $n_\infty$, necessary to calculate the saturation $s=n/n_\infty$.
Although one might try to estimate $n_\infty$ by market research,
comparison with earlier runs of the system for different goods, or
other means of educated guessing, a more pragmatic solution suggests
itself.  Namely, as in the last example in Figure~\ref{fig:incentive_examples},
setting the price to zero after some finite time, respectively at an
\textit{a priori} given $n_\infty$ obtains a natural condition for closure
of the market.

\textbf{Market closure} in this manner runs
  somewhat counter to the aim of maximising shares 
yet the 
effect on the turnover can be limited if the price becomes
small enough at late times.
Market closure yields
  the additional benefit of effectively rewarding late buyers by a
  rebate, which makes additional sense
  when looking beyond the level of a single run if the IM system.
  Then, the possible frustration of late buyers with low $v_i$ 
  might deter them from partaking
  in the IM market for a following good.   
Note
that such a procedure is not too uncommon for goods whose value is
to a large extent determined by its information content --- although
on a larger timescale than we would envisage for multi-level IM.
For instance, many academic publishers are now distributing classic 
scholarly titles for free.
Also, many legal codifications of intellectual property rights
foresee a forfeiture after a certain period.

Information is an essential tool for running a multi-level
IM system.  It is desirable to
  decouple the agents' decisions from the price and bind it more
  strongly to the incentive.  For that, a precondition is the proper
  information of the market about the expected incentive, that is,
  viable IM depends to some extent on market transparency.  
  The examples of Section~\ref{sec:cases} show that agents
  can have a rather precise estimate on their incentive using local
information.  
The operator of a multi-level IM system could support this by providing
some information of his own, but perhaps not all, since particularly the 
absolute market
size of a certain is a potentially useful information for competitors.
Namely, in a competitive situation the additional difficulty arises that 
the $s$ cannot be determined by a single party which may at best
know its own partial market size.
For instance, in order to avoid closing the market to early, close observation
of competitors prices, respectively, activity of peer-to-peer networks
distributing the good to free riders, becomes indispensable. 

Mixed forms of dynamical price settings can be envisaged\eg a positive
correlation of $\pi$ with the buying frequency, combined with a frequency or
price threshold below which the price is set to zero and the market closed.
In any case, as Section~\ref{sec:fr} and 
Section~\ref{sec:competition-sim} showed designing the optimal price schedule
is a complex task, in particular in competitive situations.
\subsection{Roots and Market Cannibalisation}\label{sec:root}
For the originator of the good,
whom we assume for the time being also to control the IM system,
there are basically two ways to extract revenues from the  market: 
He can either act as the market's \textbf{root}.
That is he is the first reseller, paying himself
a price equal to the original production cost of the
good. Or he can use a commission model
(combinations of both cases are of course thinkable).
The analysis of network effects yields an argument that the former
possibility is in principle inferior to the latter.
For the originator's revenues scale with the total revenues in the market as
$O(n_\infty)$ for large market sizes, while the revenues of a root go only with
$O(\ln n_\infty)$.
This is in essence a consequence of the fact that a pure multi-level market
cannot easily be monopolised by a single reseller, or even a group of them.
In turn it explains why commission models are a standard practise in 
multi-level marketing.

This leads us to a caveat with respect to the crucial assumption
underlying our market model, namely homogeneity.
If the market is biased in the sense that
there are groups of agents with systematically higher trading
capacities than others, this assumption breaks down.  Heuristically,
considering only an average agent in a structureless market should be
a good approximation if the number of potential participants is large
and consists of a more or less homogeneous group of individuals, 
like one with special personal preferences\eg musical.  
In practise, large music labels running direct sale web sites are 
the counterexample where this heuristics is most blatantly violated.
If such a label concurrently 
offers one of their titles through a multi-level IM system, 
the chances of the average buyer to buy from this root source are
much higher than to meet any other market participant.
The same argument exhibits the imminent danger that the market
can be cannibalised at an early stage by a player with overwhelmingly 
high communication capacity\eg a popular web site, who could
then obtain a practical monopoly.
Some studies indicate that monopoly creation could be a rather natural 
effect in e-commerce~\cite{MH03}.
While the originator of the good is not too affected by this phenomenon
if he uses a commission model, the other buyers'
incentives are always negatively affected.
To what
extent the market can be levelled by means of the IM system\eg
by providing equal communication capacities to all participants,
restricting or controlling resale volumes or frequencies, etc.,  
warrants separate discussion.
\section{Conclusions}\label{sec:conclusions}
Let us briefly note some directions for further work.
On the theoretical side it would be desirable to improve the
both the monetary flux model and the competition model to account
for\eg market inhomogeneities in the former and further externalities'
influence on the decision mechanism of the agents in the latter.
In particular, a better justified model for the multiplier effect
and a proper incorporation of other network effects is wanting.
More refined simulations of multi-level markets in the framework
of agent-based computational economics~\cite{TESF}, can be useful.
A proper treatment of multi-level markets and the competition
of goods therein from the viewpoint of theoretical economics
should also answer questions of optimality, equilibria, and their
stability.
The free-rider problem in multi-level IM should also be treated in a more theoretical
approach using the principal-agent model~\cite{LM} and aiming at describing the effect of
the incentive on the moral hazard incurred by the agents.

Pragmatically and in order to design proper market mechanisms and actual
systems for multi-level IM, the most daunting task from the present
viewpoint is to ensure equal opportunities for resellers in the market\ie
to practically corroborate the theoretical assumption of homogeneity.
\appendix
  \section{Proofs of Propositions}\label{sec:proofs}
\begin{proof}[Proof of \refP{zero-sum}]
  For $\EPS>0$ consider
  \begin{align*}
    \int_\EPS^1\dd s \int_s^1 \frac{\pi(s')}{s'}\,\dd s' &=
    \int_\EPS^1\dd s' \int_\EPS^{s'} \frac{\pi(s')}{s'}\, \dd s \\
    &=  \int_\EPS^1 \frac{s'-\EPS}{s'}\pi(s')\,\dd s' \\
    &= \int_\EPS^1 \pi(s')\,\dd s' - \EPS\int_\EPS^1 \frac{\pi(s')}{s'}\,\dd
    s'.
  \end{align*}
  If $\pi(s)$ is bounded on $[0,1]$ as assumed then the second term is
  of order $O(\EPS\ln\EPS)$ and therefore vanishes as $\EPS\searrow0$.
  The first term converges to
\[
\int_0^1\pi(s)\,\dd s,
\]
as desired.
\end{proof}
\begin{proof}[Proof of \refP{inverse}]
  For $\pi\in\mathcal{V}$, $(K\pi)(s)$ is a continuously differentiable
  function in the interval $(0,1]$ with derivative
  $-\pi(s)/s-\dot{\pi}(s)$.  The latter is of order $1/s$ as
  $s\searrow0$ since $\pi$ stays bounded at zero.  For the same reason,
  the integral in $K\pi$ is $O(\ln s)$ $(s\to0)$, which is $o(1/s)$,
  showing $(K\pi)(s)=o(1/s)$ $(s\to0)$.  $K\pi$ satisfies the zero-sum
  condition due to \refP{zero-sum}.  Thus $K\pi\in\mathcal{W}$ and we
  can apply $\check{K}$ to obtain
  \begin{align*}
    (\check K K \pi) (s) & =
    \check K \left( \int_\sigma^1 \frac{\pi(s')}{s'}\,\dd s'-\pi(\sigma)\right) \\
    & = -\frac{1}{s} \int_0^s \left( -\frac{\pi(\sigma)}{\sigma}-\dot{\pi}(\sigma) \right)\sigma\,\dd\sigma\\
    & = \frac{1}{s} \left( \int_0^s \pi(\sigma)\,\dd\sigma - \int_0^s \pi(\sigma)\,\dd\sigma
      + \bigl[ \pi(\sigma)\sigma \bigr]_0^s
    \right)\\
    &= \pi(s).
  \end{align*}
  On the other hand, if $\dot{v}_i=O(1/\sigma)$ $(\sigma\to0)$, then the last
  calculation showed that $\check{K}$ can be applied to it and obtains
  a differentiable function in $(0,1)$ which extends continuously to
  $[0,1]$.  That is $\check{K}v_i\in\mathcal{V}$ and we calculate for
  $s>0$
\begin{align*}
  (K\check{K}v_i)(s)&= -\int_s^1 \frac{1}{s'^2}\int_0^{s'}
  \sigma\dot{v}_i(\sigma)\,\dd\sigma\,\dd s'+
  \frac{1}{s}\int_0^s\sigma\dot{v}_i(\sigma)\,\dd\sigma\\
  &= -\int_s^1\dot{v}_i(s')\,\dd s' + \int_0^1\sigma\dot{v}_i(\sigma)\,\dd\sigma \\
  &= v_i(s)-v_i(1)+\lim_{\EPS\searrow0}\left(-
    \int_\EPS^1v_i(\sigma)\,\dd\sigma+\bigl[\sigma v_i(\sigma)\bigr]_\EPS^1
    +\int_0^\EPS\sigma\dot{v}_i(\sigma)\,\dd\sigma \right) \intertext{In the last
    step, we used continuity of $v_i$ at $1$.  Now, since
    $v_i=o(1/\sigma)$, $\dot{v}_i=O(1/\sigma)$ $(\sigma\to0)$, the limit can be
    assumed and yields}
  &= v_i(s)-v_i(1)+\int_0^1v_i(\sigma)\,\dd\sigma+v_i(1)\\
  &= v_i(s),
\end{align*}
where the zero-sum condition has been used.
\end{proof}
\begin{proof}[Proof of \refP{sharp-positive}]
  Partial integration yields
\[
\int_0^s\sigma \dot{v}_i(\sigma)\,\dd\sigma= \int_0^sv_i(\sigma)-sv_i(s),
\]
where we have used that $\sigma v_i(\sigma)\to0$ for $\sigma\searrow0$ if $\pi$ is
$C^1$, cf.\ \refP{inverse}.  The result follows upon inserting the
above equation into the inequality $\pi(s)>0$ and using \refP{inverse}.
\end{proof}
\begin{proof}[Proof of \refP{asymptoticbound}]
  We have for $sn_\infty$ integer
  \begin{align*}
    \ABS{v_i(s)-\overline{v}_i(sn_\infty)}&=
    \ABS{%
      \int_s^1\frac{\pi(s')}{s'}\,\dd s'
      -\sum_{k'=sn_\infty+1}^{n_\infty}\frac{\pi(k'/n_\infty)}{k'-1}%
    }\\
  &=\lim_{\EPS\searrow0}\ABS{%
    \int_s^{1+\EPS} \pi(s') \left( \frac{1}{s'}-
      \sum_{k'=sn_\infty+1}^{n_\infty}\frac{\delta(s'-k'/n_\infty)}{s'}
      \right)\,\dd s'%
    } \intertext{where we extended $\pi$ continuously in a small
      interval $[1,1+\EPS]$, and used Dirac's $\delta$-function to
      incorporate the sum in the integral.  Now, the non-negative
      factor $\pi$ can be drawn out to estimate}
    &\leq \pi_{\mathrm{max}} \ABS{ \Psi(sn_\infty) - \ln s - \Psi(n_\infty) } \\
    &= \pi_{\mathrm{max}} \bigl( \ABS{ \Psi(sn_\infty) - \ln sn_\infty} +
    \ABS{\ln n_\infty - \Psi(n_\infty)} \bigr)
  \end{align*}
  Using the asymptotic expansion of the $\Psi$ function for $r$ a
  positive integer, see~\cite[p.~295]{OLV74}, we obtain
\[
\ABS{\ln r-\Psi(r)}\leq
\frac{1}{2r}+\sum_{m=1}^n\frac{\ABS{B_{2m}}}{2mr^{2m}}, \quad\text{for }
n\geq0,
\]
where $B_{2m}$ is the $2m$-th Bernoulli number. From this follows the claim.
\end{proof}
\begin{proof}[Proof of \refP{rho}]
The assertions on monotonicity follow from positivity of probability
distributions (for $\Delta$) and stochastic dominance (for $p$, respectively, $q$).
Symmetries i) and ii), from which iii) follows directly, are easy calculations.
\end{proof}
\section{Weibull Distribution of Individual Utilities}
\label{sec:weibull-rho}
\begin{figure}[p]
  \centering \resizebox{0.6\textwidth}{!}{\includegraphics{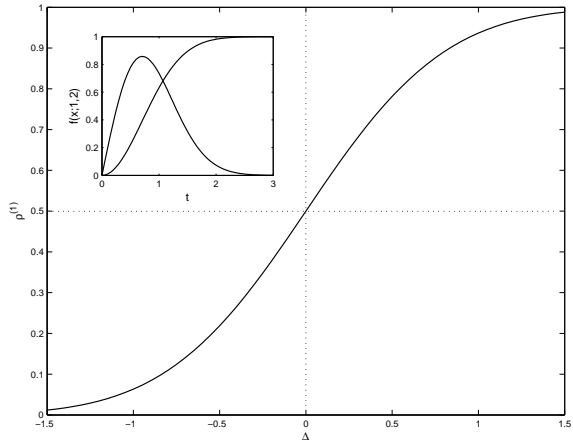}}
  \caption{%
    Decision probability $\ra(\Delta)$ resulting from popularity distributions 
    given by translates of the Weibull distribution $f(x;1,2)$.
    The PDF and CDF of $f(x;1,2)$ 
    are shown in the inlay.}
  \label{fig:rho_weibull}
\end{figure}

The Weibull distribution is widely used in reliability and lifetime estimation.
It is defined by the PDF
\[
f(x;a,b)\DEF b a^{-b} x^{b-1} \ee^{-(x/a)^b} \chi_{(0,\infty)}(x),
\] 
where $\chi_{(0,\infty)}(x)$ is the characteristic function of the positive half axis.
For $a=1$, $b=2$ it reduces to the utility distribution used in Section~\ref{sec:cases}
\[
\pdf(x)\DEF f(x;a,b)=2x\ee^{-x^2} \chi_{(0,\infty)}(x),
\]
the CDF of which is
\[
\pdf(x)=1-\ee^{-x^2}\chi_{(0,\infty)}(x).
\]
If \pdf(\pc,\uc) is given in translation form, 
formula \eqref{eq:rhocalc} yields for $\ra(\Delta)$, $\Delta\geq0$,
\begin{equation}
  \label{eq:weibull-rho}
\ra(\Delta)=1+\frac{\sqrt{2\pi}}{4}\Delta
\ee^{-\Delta^2/2}
\left(1-\erf(\Delta/ \sqrt{2})\right)
-\ee^{-\Delta^2/2}.  
\end{equation}
These functions are shown in Figure~\ref{fig:rho_weibull} where it is apparent
that \ra is concave for $\Delta>0$, in this case, and $\ra(-\Delta)=1-\ra(\Delta)$,
as follows from Proposition~\ref{prop:rho}.
%
%
\end{document}